\documentclass[a4paper,12pt]{article}

\def\half{{1\over 2}}
\def\({\left (}
\def\){\right)}
\def\[{\left [}
\def\]{\right]}

\def\la{\langle}
\def\ra{\rangle}
\def\sqr#1#2{{\vcenter{\hrule height.#2pt\hbox{\vrule width.#2pt
height#1pt \kern#1pt \vrule width.#2pt}\hrule height.#2pt}}}
\def\square{\mathchoice\sqr64\sqr64\sqr{4.2}3\sqr{3.0}3}
\def\qed{\par\noindent\rightline{$\square$}}
\def\jobo{\phantom{\tiny{\lowercase\expandafter{\jobname}.tex \ \ \today}}}
\def\job{\tiny{\lowercase\expandafter{\jobname}.tex \ \ \today}}
\usepackage[dvips]{graphicx}
\makeatletter
\renewcommand{\section}{\@startsection
{section} {1} {0mm} {-\baselineskip} {0.1\baselineskip} {\normalfont\normalsize\bfseries}}
\renewcommand{\paragraph}{\@startsection%
{paragraph}%
{4}%
{5mm}%
{-0.1\baselineskip}%
{0\baselineskip}%
{\normalfont\normalsize\rmfamily}}%
\makeatother

\makeatletter \@addtoreset{equation}{section} \makeatother
\font\BB=msbm10
\def\RR{\hbox{\BB R}}

\def\CC{\hbox{\BB C}}

\def\PP{\hbox{\BB P}}

\def\EE{\hbox{\BB E}}
\def\ZZ{\hbox{\BB Z}}

\begin{document}
%\fbox{{\bfseries REVISED}}
%\vskip 0.5truecm
\centerline{\bf THE SPECTRUM OF A MAGNETIC SCHR\"ODINGER OPERATOR } \centerline{\bf WITH
RANDOMLY LOCATED DELTA IMPURITIES} \vskip 0.25truecm \centerline{ {\bf \quad J.V.
Pul\'e}\footnote{\sl e-mail: joe.pule@ucd.ie}\footnote{\sl Research Associate, School of
Theoretical Physics, Dublin Institute for Advanced Studies.} {\bf and M.
Scrowston}\footnote{\sl e-mail: mark.scrowston@ucd.ie} } \vskip 0.5truecm
\centerline{Department of Mathematical Physics} \centerline{National University of Ireland,
Dublin} \centerline{Belfield, Dublin 4} \vskip 0.5truecm \centerline{\bfseries Abstract} We
consider a single band approximation to the random Schr\"odinger operator in an external
magnetic field. The spectrum of such an operator has been characterized in the case where
delta impurities are located on the sites of a lattice. In this paper we generalize these
results by letting the delta impurities have random positions as well as strengths; they are
located in squares of a lattice with a general bounded distribution. We characterize the
entire spectrum of this operator when the magnetic field is sufficiently high. We show that
the whole spectrum is pure point, the energy coinciding with the first Landau level is
infinitely degenerate and that the eigenfunctions corresponding to other Landau band energies
are exponentially localized.\\ \centerline{\bf Running Title: Magnetic Schr\"odinger Operator
with Delta impurities} \centerline{PACS:\enskip 71.55.J, 73.40.H, 71.45}
\newpage

\section{Introduction}
\paragraph
In recent years there has been considerable activity in the study of random magnetic
Schr\"odinger operators mainly due to their relation with the theory of the Integer Quantum
Hall Effect (IQHE). Some of these studies have incorporated the randomness into the magnetic
field$^1$, whereas others have added a random potential to the usual Landau Hamiltonian.
Without any disorder the Landau Hamiltonian has a spectrum of evenly spaced {\sl Landau
levels}, each one of which is an infinitely degenerate eigenenergy. When a random potential is
added these Landau levels broaden into bands. In several models$^{2-4}$ it has been shown that
for large magnetic field the spectrum at the edges of the bands is pure point, with each
eigenenergy corresponding to an exponentially localized state. The proofs rely on von Dreifus
and Klein's$^5$ refined version of the earlier multiscale analysis by Fr\"ohlich and
Spencer$^6$ and on percolation theory. These results are not sufficient to provide a complete
understanding of the IQHE however, as the nature of the spectrum in the interior of the band
is crucial in explaining the observed plateaux$^7$. In one special case$^{8,9}$ the spectrum
has been completely characterized. In this work the authors consider a random potential
consisting of zero-range scatterers (delta functions) situated on the sites of a regular
lattice. In the first paper$^8$, they show that, in the case of a single-band approximation,
the whole spectrum is pure point, with exponentially localized states for all energies except
the original Landau level. They prove also that this level remains infinitely degenerate.
These results are improved in a later work$^9$ where they obtain similar results for the
unprojected Hamiltonian. They adopt a simple proof of localization by Aizenman and
Molchanov$^{10}$ which utilizes low moments of the resolvent kernel.
\paragraph
The purpose of this work is to generalize the above for the case of a magnetic Schr\"odinger
operator with randomly distributed delta impurities. Specifically, the random potential
consists of point scatterers, delta functions, positioned in  unit squares which are centered
on the Gaussian integers, so that it is possible to have up to four scatterers arbitrarily
close together. The strengths of the scatterers will also be random. We consider a
two-dimensional infinite system of noninteracting electrons moving in a uniform magnetic field
of strength $B$ and the random potential $V$. The precise hypotheses on the probability
distributions will be stated in Section II.

In the symmetric gauge the vector potential is given by ${\bf A}({\bf r})=\half({\bf r}\times
{\bf B})$ and the Hamiltonian is
\begin{equation}\label{opb:fullham}
H=(-i\nabla -{\bf A}({\bf r}))^2+V({\bf r}).
\end{equation}
When the magnetic field is sufficiently strong in comparison to the random potential the
Landau bands do not overlap and it is sufficient to consider the projection of the Hamiltonian
onto only one of them. The Hamiltonian retricted to the $n$th level is
\begin{equation}\label{opb:projham}
H_n=B(2n+1)P_n+P_nVP_n,
\end{equation}
where $P_n$ denotes the projection onto the level. The first term comes from the decomposition
of the purely kinetic part of (\ref{opb:fullham}) and can be dropped as it modifies the energy
only by a constant. Note that the resulting Hamiltonian is a random integral operator instead
of a differential operator and that the kernels of $P_n$ are known explicitly. For simplicity,
in this paper, we restrict ourselves to the case $n=0$ but the case $n\neq 0$ can be treated
similarly.
\paragraph
For our model, in the special case where the support of the positional probability
distribution is bounded within each unit box so that a corridor exists between impurities, the
method of Aizenman and Molchanov yields a simple proof of localization$^{11}$. However, for
general distributions of position, impurities can become arbitrarily close to each other and
we are not able to use their method. This is partly due to possible resonances; that when
impurities can become arbitrarily close together the low moments of the resolvent kernel do
not converge rapidly enough. In this paper we use the modification of the Theorem of von
Dreifus and Klein given in Ref. 2 to show exponential localization of states corresponding to
each of the  eigenenergies separately (except the original Landau level eigenenergy). We do
this by studying, at fixed energy, the behaviour of the generalized eigenfunctions at the
impurity sites only, thus reducing the problem to the study of a random matrix. The
eigenfunctions of this matrix are related to the eigenfunctions of the Hamiltonian in such a
way that exponential decay of the former implies exponential decay of the latter. Then using
Kotani's `trick' $^{12}$ we can show exponential decay for all eigenenergies in an interval
with probability one implying that the whole spectrum is pure-point. That the original Landau
level eigenenergy remains infinitely degenerate has been shown in Ref. 13 for the case of a
Poisson distribution of impurities. The result given here is similar and so only a sketch of
the proof is given.

The paper is organised as follows. In Section II we give a precise definition of the model. In
Section III we characterize the spectrum as a set, state the main theorem and show infinite
degeneracy of the original eigenenergy. Also in this section we relate the Hamiltonian to a
lattice operator and state our version of the adapted von Dreifus-Klein Theorem. Section IV
contains the main work of this paper, where the conditions of the main theorem are checked. In
Section V we use Kotani's trick to show exponential decay and pure-point spectrum with
probability one.

\section{Definition and Boundedness of the Hamiltonian}
\paragraph
Let $\omega_n$, $n\in\ZZ[i]\equiv\{ n_1 +in_2 :\ (n_1,n_2)\in\ZZ^2\}$, the Gaussian integers,
be independent, identically distributed (i.i.d) random variables representing the strengths of
the impurities. We shall assume that their distribution is given by an absolutely continuous
probability measure $\mu$ whose support is a compact interval $X=[a,b]\subset\RR$ containing
the origin and whose density $\rho $ is bounded by a constant $\rho_b$. We let $\Omega_1
=X^{\ZZ[i]}$ and $\PP_1 =\prod_{n\in\ZZ[i]}\mu$.

Define the unit squares centred at $n\in\ZZ [i]$: $$ B_n =\{ z\in\RR^2\,\vert\, n_i-\half\le
z_i < n_i+\half\,,\,n\in\ZZ [i]\,,\,i=1,2\,\}. $$
Let $\zeta_n =n+ {\tilde \zeta}_n$, $n\in\ZZ[i]$, represent the positions of the impurities in
the complex plane. ${\tilde \zeta }_n$, $n\in\ZZ [i]$ are i.i.d. random variables. We shall
assume that their distribution is given by a probability measure $\nu$ with support equal to
$B_0$ and density $r $ bounded by a constant $r_b$. We let $\Omega_2 =\times_{n\in\ZZ[i]}B_0$
and $\PP_2 =\prod_{n\in\ZZ[i]}\nu$. Our probability space will be $\Omega
=\Omega_1\times\Omega_2$ with probability measure $\PP =\PP_1\times\PP_2$.

For $m\in\ZZ [i]$ let $\tau_m$ be the measure preserving automorphism of $\Omega$
corresponding to translation by $m$:
\begin{equation}
(\tau_m(\omega,\zeta))_n=(\omega_{n-m},\zeta_{n-m})\ .
\end{equation}
The group $\{\tau_m:m\in\ZZ [i]\}$ is ergodic for the probability measure $\PP$.

Let ${\cal H}=L^2(\CC)$ and let ${\cal H}_0$ be the eigenspace corresponding to the lowest
eigenvalue (first  Landau level) of the kinetic part of the Hamiltonian defined in
(\ref{opb:fullham}) and let $P_0$ be the orthogonal projection onto ${\cal H}_0$. The
Hamiltonian for our model is the operator on ${\cal H}_0$ given formally by
\begin{equation}
H\(\omega,\zeta\) = {\pi\over {2\kappa} } P_0 V\( \cdot , (\omega,\zeta)\) =  {\pi\over
{2\kappa} }P_0 V\( \cdot , (\omega,\zeta)\)P_0
\end{equation}
where $(\omega,\zeta) \in \Omega$ and
\begin{equation}
V\(z, (\omega, \zeta)\) = \sum_{n\in\ZZ [i]} \omega_n \delta(z-\zeta_n).
\end{equation}
Note that $H$ coincides with $H_0$ in (\ref{opb:projham}) up to the term $BP_0$ and a
multiplicative constant and that the lowest Landau energy is now shifted to zero. The
projection $P_0$ is an integral operator with kernel
\begin{equation}\label{opb:kernelproj}
P_0 \(z,z'\) = {{2\kappa} \over { \pi}} \exp [- {\kappa} \vert z - z' \vert^2 - 2i \kappa z
\wedge z'],
\end{equation}
where $\kappa=B/4$ and $ z  \wedge z'={\cal R}z{\cal I}z'-{\cal I}z{\cal R}z'$, ${\cal R}z$
and ${\cal  I}z$ being the real and imaginary parts of $z$ respectively. Note that if $\psi
\in {\cal H}$ then $\psi \in {\cal H}_0$ if and only if $\psi(z) = f(z) e^{- \kappa \vert z
\vert^2}$ where $f(z)$ is entire. Using (\ref{opb:kernelproj}) we can write the Hamiltonian in
the form $$ H = \sum_{n\in\ZZ [i]} \omega_n f_{\zeta_n} \otimes {\overline{f_{\zeta_n}}}, $$
where for $\zeta \in \CC$,
\begin{equation}
f_\zeta (z) =  \sqrt{{\pi \over {2 \kappa}}}P_0 (z, \zeta) = \sqrt{{2\kappa} \over { \pi}}
\exp [2\kappa {\bar \zeta}z- {\kappa}( \vert \zeta \vert^2 + \vert z \vert^2)].
\end{equation}
Note that $\Vert f_\zeta \Vert =1$, $\la f_\zeta\,,\ f_{\zeta'}\ra
=\sqrt{{\pi\over{2\kappa}}}f_{\zeta'}(\zeta)$ and that $H$ is an integral operator with kernel
\begin{equation}
H(z,z') = \sum_{n\in\ZZ [i]} \omega_n f_{\zeta_n} (z) {\overline{f_{\zeta_n} (z')}}.
\end{equation}
We first obtain a bound on $H(z,z^\prime )$ which implies that $H$ is bounded. We give the
following simple estimate without proof.

\noindent {\bf Lemma 2.1}:\ {\sl For $s,\ t > 0$ and $z,\ z'\in \CC$ $$ \sum_{n \in \ZZ[i]}
e^{-s \vert z -\zeta_n \vert^2} e^{-t \vert\zeta_n - z' \vert^2} \leq K(s+t) e^{- {{st} \over
{s + t}} \vert z - z' \vert^2}. $$ where $$ K(s)= 9+8e^{-s} + 4\left( {\pi\over
{s}}\right)^\half +{4\over s} . $$ }
The above Lemma implies that $\vert H(z,z')\vert$ is bounded above by
\begin{equation}
{{2R\kappa}\over \pi} K(2\kappa)e^{- {\kappa \over 2} \vert z - z' \vert^2},
\end{equation}
where $R=\max(\vert a\vert, \vert b \vert)$. Therefore $H$ is bounded and
\begin{equation}
\Vert H \Vert \leq 4R K(2\kappa).
\end{equation}
Note that the heat kernel is $$ P_t(z,z')= {1\over{2\pi t}} e^{- {1 \over {2t}} \vert z - z'
\vert^2} $$ and the corresponding operator has unit norm. From now on we take $\kappa$
sufficiently large so that $K(2\kappa )<10$ and we let ${\bar R}=40R$ so that $\Vert H\Vert\le
{\bar R}$.

\section{The Spectrum of $H$.}
\paragraph
Let $\{U_z:z\in\CC\}$ be the family of unitary operators on ${\cal H}$ corresponding to the
magnetic translations: $$ (U_zf)(z')= e^{2i\kappa z\wedge z'}f(z+z'). $$ Then for $m\in\ZZ
[i]$
\begin{equation}\label{opb:magtrans}
U_mH(\omega,\zeta)U_m^{-1}=H(\tau_m(\omega,\zeta)).
\end{equation}
Note that $[P_0,U_z]=0$ for all $z\in\CC$ so that $U_z{\cal H}_0\subset {\cal H}_0$. Also
$U_{z_1}U_{z_2} =e^{2i\kappa z_2\wedge z_1}U_{z_1+z_2}$. The ergodicity of $\{\tau_m:m\in\ZZ
[i]\}$ and equation (\ref{opb:magtrans}) together imply that the spectrum of $H(\omega,\zeta)$
and its components are non random (see Ref. 14 Th V.2.4).

{\bf Lemma 3.1}:\ {\sl With probability one} $$ [4a,4b]\subset\sigma \(H(\omega,\zeta)\). $$
{\bf Proof: } It is sufficient to prove that for each $E\in [4a,4b]$ and for all $\delta>0$,
there exists $\Omega'\subset\Omega$ with $\PP (\Omega')>0$ and $\psi\in {\cal H}_0$ with
$||\psi||=1$ such that for all $(\omega,\zeta)\in\Omega'$, $||\(H(\omega,\
\zeta)-E\)\psi||<\delta.$ \vskip .2cm \noindent Let $B =\{0,1,i,1+i\}$. Choose $E\in [4a,4b]$
and $D$ such that $\sum_{n:\vert\zeta_n\vert\ge D}e^{- \kappa\vert\zeta_n
-\zeta_0\vert^2}<\delta /4R$, where $R=\max (\vert a\vert ,\vert b\vert )$, and let $$
\Omega_2^\prime =\{\zeta\in\Omega_2 :\vert\zeta_n -\zeta_0\vert\le {\delta\over
{4E\sqrt{2\kappa}}}\ \forall\, n\in B \} $$ then since for all $n\in B$, the impurities
$\zeta_n$ and $\zeta_0$ can be arbitrarily close to one another, $\PP (\Omega_2^\prime )>0$.
Let $$ \Omega_1^\prime =\{\omega\in\Omega_1 :\vert\omega_n -{E\over 4}\vert <{\delta\over 16}\
\forall\, n\in B {\hbox {\rm { and }}} \max_{m\notin B\,:\vert\zeta_m\vert<D}\vert\omega_m
\vert K(\kappa )<{\delta\over 4}\}. $$ Since $E/4$ and $0$ are in the support of $\mu$, $\PP
(\Omega_1^\prime )>0$. Let $\Omega^\prime =\Omega_1^\prime\times\Omega_2^\prime$, then $\PP
(\Omega^\prime )>0$.

Now
\begin{eqnarray*}
(Hf_{\zeta_0}-Ef_{\zeta_0})(z) &=&\sum_{n\in B}(\omega_n -{E\over 4}) \la
f_{\zeta_n},f_{\zeta_0} \ra f_{\zeta_n}(z)+E\left( {1\over 4}\sum_{n\in B} \la
f_{\zeta_n},f_{\zeta_0} \ra f_{\zeta_n}(z)-f_{\zeta_0}(z)\right)\nonumber\\ &+&\sum_{n\notin
B:\vert\zeta_n\vert <D}\omega_n \la f_{\zeta_n}, f_{\zeta_0} \ra
f_{\zeta_n}(z)+\sum_{n:\vert\zeta_n\vert\ge D} \omega_n \la f_{\zeta_n},f_{\zeta_0}  \ra
f_{\zeta_n}(z).
\end{eqnarray*}
Hence
\begin{eqnarray*}
\Vert Hf_{\zeta_0}-Ef_{\zeta_0} \Vert &\le &{\delta\over 4}+{E\over 4} \Vert \sum_{n\in B} \la
f_{\zeta_n},f_{\zeta_0} \ra f_{\zeta_n}-4f_{\zeta_0}\Vert\nonumber\\ &+&\sum_{n\notin
B\,:\vert\zeta_n\vert <D}\vert\omega_n\vert\, \vert\la f_{\zeta_n},f_{\zeta_0}\ra \vert
+R\sum_{n:\vert\zeta_n\vert\ge D} e^{-\kappa\vert\zeta_n -\zeta_0\vert^2}.
\end{eqnarray*}
It is easily seen that $\Vert \la f_{\zeta_n},f_{\zeta_0}\ra f_{\zeta_n}-f_{\zeta_0} \Vert^2
=1-\vert  \la f_{\zeta_n},f_{\zeta_0}\ra \vert^2$ is bounded by $2\kappa\vert\zeta_n
-\zeta_0\vert^2$, and therefore for all $(\omega ,\zeta )\in\Omega^\prime$,
$$ \Vert Hf_{\zeta_0}-Ef_{\zeta_0} \Vert < \delta. $$ \qed

We now state the main theorem, which we will prove in the sequel.

{\bf Theorem 3.2 } {\sl There exists $\kappa_0>0$ such that for $\kappa>\kappa_0$, with
probability one,

(a) $0$ is an eigenvalue of $H$ with infinite multiplicity,

(b)\ $\sigma_{\hbox {\rm cont }}(H) =\emptyset$,

(c) if $\lambda\in\sigma(H)\backslash\{ 0\}$, is an eigenvalue of $H$ with eigenfunction
$\psi$, then $\psi$ decays exponentially with rate  $\ge \kappa^{1/4}$. }

We will start by showing part (a). The lemma is very close to results proven in Refs 8 and 13
so only a sketch of the proof will be given.

{\bf Lemma 3.3}:\  {\sl There exists $\kappa_1
>0$ such that for $\kappa >\kappa_1$, with probability one, 0 is an eigenvalue of $H$ with
infinite multiplicity.}
\newline
{\bf Proof: } Let $$ \psi_\zeta (z) = \prod_{n\in\ZZ [i]} (1 - {z \over \zeta_n}) e^{{z\over
\zeta_n} +{{z^2}\over {2\zeta_n^2}}}. $$ If we can show that the sums $\sum_n
1/\vert\zeta_n\vert^3$ and $\sum_n 1/\zeta_n^2$ converge independently of $\zeta$ then it
follows from the theory of entire functions (see Ref. 15) that there exists $A>0$ and $R>0$,
both independent of $\zeta$ such that for $\vert z\vert>R$, $\vert \psi_\zeta (z)\vert \leq
e^{A\vert z \vert^2}$. The first sum is easily bounded, the second can be bounded by utilising
the four-fold rotational symmetry of the $\ZZ [i]$ to cancel any large contributions. Let
$\phi_k (z) =  z^k \psi_\zeta (z) e^{-\kappa \vert z \vert^2}$ for $k \geq 1$, then if $\kappa
>A$, the $\phi_k$'s are in ${\cal H}_0$ and $\phi_k (\zeta_n)=0$ for all $n\in\ZZ [i]$.
Therefore $H\phi_k =0$ for all $k\ge 1$. Moreover if $\sum^N_{j=1} a_{k_j} \phi_{k_j} = 0$
then $\sum^N_{j=1} a_{k_j} z^{k_j} = 0$ for $ z \notin \{ \zeta_n \}$. Therefore $\sum^N_{j=1}
a_{k_j} z^{k_j} \equiv 0$ and thus the $ a_{k_j}$'s are zero implying that the $\phi_k$'s are
independent.\qed

For Theorem 3.2 parts (b), (c) we can simplify the problem by studying the behaviour of the
generalized eigenfunctions at the impurity sites only. We have
$$ (H\psi )(z)={\pi\over {2\kappa}}\sum_{n\in\ZZ [i]}\omega_n P_0 (z,\zeta_n )\psi (\zeta_n )
$$
and thus if $(H\psi )(z)=\lambda\psi (z)$,
\begin{equation}\label{opb:sitetise}
{\pi\over {2\kappa}}\sum_{n\in\ZZ [i]}\omega_n P_0 (z,\zeta_n )\psi (\zeta_n )=\lambda\psi (z)
\end{equation}
which evaluated at $\zeta_m$ gives
$$ {\pi\over {2\kappa}}\sum_{n\in\ZZ [i]}\omega_n P_0 (\zeta_m ,\zeta_n ) \psi (\zeta_n
)=\lambda \psi (\zeta_m ) $$
Let $\omega_n \psi (\zeta_n )=\xi_n$. Then
\begin{equation}
{\pi\over {2\kappa}}\sum_{n\in\ZZ [i]} P_0 (\zeta_m ,\zeta_n ) \xi_n ={\lambda\over
{\omega_m}}\xi_m
\end{equation}
We can thus reduce the problem to the study of a random matrix which has $\omega$-dependent
elements on the diagonal and $\zeta$-dependent rapidly decaying off-diagonal elements. We
write this matrix as a sum of a diagonal matrix and an off-diagonal matrix as defined below.

Let ${\cal M}=l^2 (\ZZ [i]) $. Define the operators $M_0$, and $V_{\omega}^{\lambda}$ on
${\cal M}$ as follows.
\begin{eqnarray}\label{opb:defnM}
& &\langle m\vert M_0\vert n\rangle ={\pi\over 2\kappa}P_0(\zeta_m
,\zeta_n)(1-\delta_{mn})\nonumber\\ & &\langle m\vert V_{\omega}^{\lambda}\vert n\rangle
=\left( 1-{\lambda\over \omega_n}\right) \delta_{mn}.
\end{eqnarray}
For a proof of Theorem 3.2 part (c) we note that the eigenvectors $\xi$ of
$M^{\lambda}=M_0+V_{\omega}^{\lambda}$, are related by an explicit formula to the generalized
eigenfunctions $\psi$ of $H$ in such a way that exponential decay of the former implies
exponential decay of the latter:

>From (\ref{opb:sitetise}), if $\lambda\ne 0$
$$ \psi (z)={\pi\over {2\kappa\lambda}}\sum_{n\in\ZZ [i]} P_0(z,\zeta_n )\xi_n $$
If $\xi_n $ decays exponentially, $\vert\xi_n \vert\le Ce^{-m\vert\zeta_n \vert }$ we have
\begin{eqnarray}\label{opb:expimpliesexp}
\vert\psi (z)\vert &\le &{C\over\lambda}\sum_{n\in\ZZ [i]} e^{-\kappa\vert
z-\zeta_n\vert^2}e^{-m\vert\zeta_n \vert} \le {C\over\lambda }e^{-m\vert z\vert}\sum_{n\in\ZZ
[i]} e^{-\kappa\vert z-\zeta_n\vert^2}e^{m\vert z-\zeta_n \vert}\nonumber\\ &\le &
{C\over\lambda}e^{-m\vert z\vert} e^{{m^2}\over {2\kappa }} \sum_{n\in\ZZ [i]}
e^{-{\kappa\over 2}\vert z-\zeta_n\vert^2} \le {C\over\lambda}e^{{m^2}\over {2\kappa
}}K({\kappa\over 2}) e^{-m\vert z\vert}
\end{eqnarray}
and $\psi (z)$ decays exponentially, where we have bounded the sum by taking $s=\kappa /2$,
$t=0$ in Lemma 2.1.

Thus we want to show that the eigenvectors for the eigenvalue equation $M^{\lambda}\xi =0$
decay exponentially for $\lambda\ne 0$. We will do this by the same method as in Ref. 8. First
we will need to make a few definitions.

For regions $\Lambda \subset \ZZ[i]$ we define $M^\lambda_\Lambda$ to be the restriction of
$M^\lambda $ to $l^2(\Lambda)$. If $E \notin \sigma (M^\lambda_\Lambda)$ then the Green
function
\begin{equation}
\Gamma^\lambda_\Lambda(E) = (M^\lambda_\Lambda - E)^{-1}
\end{equation}
is well-defined. In particular, we shall consider the regions
\begin{equation}
\Lambda_L (n) = \{ n' \in \ZZ[i] \vert\ :\ \vert n'-n \vert_\infty <L/2 \}
\end{equation}
for $n \in \ZZ[i]$ and $L > 0$.

{\bf Definition. }{\it Fix constants $\beta \in (0,1)$ and $s \in ({1\over 2},1)$. Given a
configuration $(\omega ,\zeta )$, a square $\Lambda_L (n)$ is called $(m,E)$-regular for some
$m>0$ and $E \in \RR$ if the following two conditions are satisfied:
$$ d(E,\sigma(M^\lambda_{\Lambda_L(n)}(\omega ,\zeta )))
>\half e^{-L^\beta}, \leqno (RA)
$$
$$ \vert\langle n \vert \Gamma^\lambda_{\Lambda_L(n)}(E)\vert n'\rangle\vert\leq e^{-mL}
\leqno (RB) $$
for all $n'\in \tilde \Lambda_L (n) $ where $\tilde \Lambda_L (n) = \Lambda_L(n) \setminus
\Lambda_{\tilde L} (n) $ with $\tilde L = L - L^s$. $\Lambda_L (n)$ is called singular if it
is not regular.}

We now state a theorem which is an variant of the main theorem in Ref. 2 where the von Dreifus
and Klein Theorem is adapted from the case of a tight-binding (finite range) Hamiltonian to
the case where the Hamiltonian has a long range hopping term with Gaussian decay. It states
conditions under which the eigenvectors of the random matrix $M^\lambda$ with eigenvalue $0$
decay exponentially.

{\bf Theorem 3.4} {\it  Fix constants $\beta \in (0,1)$, $s \in ({1\over 2},1)$, $\gamma\in
(0,1)$, $p > 2$, $q > 4p+12$. There exists $Q_0 > 0$ depending on all these constants but {\bf
independent of $\lambda$ and $\kappa >1$} such that the following holds: If for
$\lambda,\,\kappa$ the conditions {\rm (P1)} and {\rm (P2)} are satisfied, where

\begin{description}
\item[\hspace{0.5cm}] {\rm (P1)} There exists an $L_0 > Q_0$ and $m_0 $
such that
\begin{equation}
\PP \left\{ \Lambda_{L_0} (0) \hbox{ is } (m_0,0)\hbox{-regular} \right\} \geq 1-L_0^{-p}
\end{equation}
\item[\hspace{0.5cm}] {\rm (P2)} There exists $\eta >0$ such that, for
all $E \in (-\eta,\eta)$ and for all $L > L_0$,
\begin{equation}
\PP \left\{ d\left(E , \sigma \left( M^\lambda_{\Lambda_L (0)} \right) \right) < e^{-L^\beta}
\right\} < L^{-q},
\end{equation}
\end{description}
\noindent then, for all $m \in (0,m_0)$, there exists $\delta>0$ depending on
$m,m_0,L_0,\beta$ and $\eta$ such that for all $E\in (-\delta,\delta)$ the eigenvectors of
$M^\lambda$ with eigenvalue $E$ decay exponentially with rate $\geq m$.}

The main work of this paper consists in proving that the conditions (P1) and (P2) are
satisfied. The conditions can be seen to consist of two types of estimate. $(RB)$ of (P1) is
an estimate of the decay of the Greens function $\Gamma^\lambda_{\Lambda_L}(0)$, while $(RA)$
of (P1) and (P2) are Wegner type estimates that require small gaps in the $\Lambda_L$
dependent spectrum. It is unusual that it is the latter estimates that will require the finer
analysis; previous works have found the decay of the Green's function to require the more
delicate study. This is because we want to show that the eigenfunctions are exponentially
decaying for arbitrarily small $\lambda$. Inspecting (\ref{opb:defnM}) we see that for
$\lambda$ small the $\omega$-dependence becomes less significant and does not give sufficient
randomness for Wegner type estimates. Therefore we have to utilise randomness provided by the
positions of the $\zeta_j$'s.

\section{Proof of the Conditions (P1) and (P2)}
\paragraph
We will begin by showing $(RB)$ of (P1). We will need the following two probabilistic lemmas
in which we fix $u>3$.

{\bf Lemma 4.1}: {\sl There exists $Q_1$ such that
\begin{equation}
\PP\left(\vert\zeta_n -\zeta_{n'}\vert>{2\over L^u} {\hbox { for all }} n,n'\in\Lambda_L,\
n\ne n' \right) \ge 1-{1\over L^{u-3}}
\end{equation}
for all $L>Q_1$. }

{\bf Proof}: The $\zeta_n$'s have a distribution with a density bounded by $r_b$ for each
$B_n$ and thus, $$ \PP\left(\vert\zeta_n -\zeta_{n'}\vert>{2\over L^u} {\hbox { for all }}
n,n'\in\Lambda_L,\ n\ne n' \right) \ge\left(1-{{4r_b}\over L^u}\right)^{L^2}. $$ By taking $L$
sufficiently large we get the result.\qed

{\bf Lemma 4.2}: {\sl There exists $Q_2$ such that
\begin{equation}
\PP\left(\left\vert 1-{\lambda\over\omega_n}\right\vert >{1\over L^u} \ \ \forall\
n\in\Lambda_L\ \right)> 1-{1\over L^{u-3}}
\end{equation}
for all $L>Q_2$. }

{\bf Proof}: Now $\displaystyle {\left\vert 1-{\lambda\over\omega_n}\right\vert\le {1\over
L^u}}$ gives us that $\displaystyle {-{1\over L^u}\le 1-{\lambda\over\omega_n} \le {1\over
L^u}}$ which implies \newline that  $\displaystyle {1+{1\over L^u}\ge {\lambda\over\omega_n}
\ge 1-{1\over L^u}}$.

Thus $\omega_n$ must fall between the bounds,
\begin{equation}
{{\vert\lambda\vert}\over {1+{1\over L^u}}} \le\ \vert\omega_n\vert\
\le{{\vert\lambda\vert}\over {1-{1\over L^u}}} .
\end{equation}
Hence
\begin{eqnarray*}
\PP\left(\left\vert 1-{\lambda\over\omega_n}\right\vert\le {1\over L^u} \right)&\le &2\rho_b
\vert\lambda\vert\left({1\over {1- {1\over L^u}}}- {1\over {1+ {1\over L^u}}}\right)
\nonumber\\ &=&{{4\rho_b\vert\lambda\vert}\over L^u}(1-{1\over L^{2u}})^{-1} \le
{{2^{u+2}\rho_b {\bar R}}\over L^u}.
\end{eqnarray*}
if $L>2$, where we have used $1-1/L^{2u}>1/2^u$. Therefore $$ \PP\left(\left\vert
1-{\lambda\over\omega_n}\right\vert >{1\over L^u} \ \ \forall\ n\in\Lambda_L\ \right)>\left(
1- {{2^{u+2} \rho_b {\bar R}}\over L^u} \right)^{(L+1)^2}. $$
By taking $L$ sufficiently large we get the result.\qed

The following Lemma is proved in Ref. 8.

{\bf Lemma 4.3}:\ {\sl  For all $\gamma\in (0,1)$, there exists $C_0(\gamma)> 0$ such that for
$\alpha>1$}
\begin{equation}
\sum_{m \in \ZZ[i]} e^{- \alpha \{ \vert z - m \vert^\gamma + \vert z' - m \vert^\gamma\}}
\leq C_0(\gamma) e^{- \alpha \vert z - z' \vert^\gamma}.
\end{equation}

The following Lemma will be used to show $(RB)$ of (P1).

{\bf Lemma 4.4}: {\sl For all $\gamma\in (0,1)$ and $u>3$, there exists $Q_3$ such that for
all $L> Q_3$ and all $\kappa >L^{4u}/4$ we have for any $n,n'\in\Lambda_L$,
\begin{equation}
\PP\(\vert\la n\vert\Gamma_{\Lambda_L}^{\lambda}(0)\vert n'\ra\vert\le 2L^u
e^{-{\kappa^{1/2}\over 8}\vert n-n'\vert^{\gamma}}\)>1-{2\over L^{u-3}} .
\end{equation}
} {\bf Proof}: In the following we let $\gamma\in (0,1)$. Using $\vert a+b\vert^2\le 2(\vert
a\vert^2+\vert b\vert^2 )$ we have $\kappa\vert\zeta_n -\zeta_{n'}\vert^2\ge\kappa(\half\vert
n-n'\vert^2 -2)
>{\kappa\over 4}\vert n-n'\vert^2>{{\kappa^{1/2}}\over 4}
\vert n-n'\vert^{\gamma}$ for $\vert n-n'\vert\ge 3$. Suppose that
$\vert\zeta_n-\zeta_{n'}\vert>2/L^u$ for all $n,n'\in\Lambda_L$, $n\ne n'$. Then for $\vert
n-n'\vert <3$  we have that $\kappa\vert\zeta_n -\zeta_{n'}\vert^2> {{\kappa^{1/2}}\over
4}\vert n-n'\vert^{\gamma}$ if ${{2\kappa^{1/2}}\over L^{2u}}> 1$. Thus we can write
$$ e^{-\kappa\vert\zeta_n -\zeta_{n'}\vert^2} \le e^{-{\kappa^{1/2}\over 4}\vert n
-n'\vert^{\gamma}} $$ and consequently $$ \vert\la n\vert M_{\Lambda_L}^0\vert n'\ra\vert \le
e^{-{\kappa^{1/2}\over 4}\vert n -n'\vert^{\gamma}}(1-\delta_{n n'}). $$
If $\vert 1-\lambda/\omega_n\vert>1/L^u$ for all $n\in\Lambda_L$ then for all
$n,n'\in\Lambda_L$ we also have, $$ \vert\la n\vert (V_{\Lambda_L}^{\lambda} )^{-1}\vert
n'\ra\vert\le L^{u} \delta_{n n'}. $$ Therefore we can write
\begin{eqnarray*}
\vert\la n\vert (V_{\Lambda}^{\lambda})^{-1} M_{\Lambda_L}^0\vert n'\ra\vert\, &&
\le\,\sum_{p\in\Lambda}\vert\la n\vert (V_{\Lambda}^{\lambda})^{-1}\vert p\ra\vert\vert\la
p\vert M_{\Lambda_L}^0\vert n'\ra\vert\\ &&=\,\vert\la n\vert
(V_{\Lambda}^{\lambda})^{-1}\vert n\ra\vert\vert\la n\vert M_{\Lambda_L}^0\vert n'\ra\vert\\
&&\le\, L^u e^{-{\kappa^{1/2}\over 8}} e^{-{\kappa^{1/2}\over 8}\vert n -n'\vert^{\gamma}}
(1-\delta_{nn'}),
\end{eqnarray*}
and
\begin{eqnarray*}
\vert\la n\vert\left( (V_{\Lambda_L}^{\lambda})^{-1} M_{\Lambda_L}^0 \right)^2\vert n'\ra\vert
&&\le\ \sum_{r\in\Lambda}\vert\la n\vert (V_{\Lambda_L}^{\lambda})^{-1}\vert n\ra\vert\vert\la
n\vert M_{\Lambda_L}^0\vert r\ra\vert \vert\la r\vert (V_{\Lambda_L}^{\lambda})^{-1}\vert
r\ra\vert\vert\la r\vert M_{\Lambda_L}^0\vert n'\ra\vert\\ &&\le\ L^{2u}e^{-{\kappa^{1/2}\over
4}} \sum_{r\in\ZZ [i]}e^{-{\kappa^{1/2}\over 8}\vert n-r\vert^{\gamma}} e^{-{\kappa^{1/2}\over
8}\vert r-n'\vert^{\gamma}} (1-\delta_{nr})(1-\delta_{rn'})\\ &&\le\
C_0(\gamma)L^{2u}e^{-{\kappa^{1/2}\over 4}} e^{-{\kappa^{1/2}\over 8}\vert
n-n'\vert^{\gamma}}.
\end{eqnarray*}
Similarly, $$ \vert\la n\vert\left( (V_{\Lambda_L}^{\lambda})^{-1} M_{\Lambda_L}^0\right)^k
\vert n'\ra\vert \le C_0(\gamma)^{k-1}L^{ku}e^{-{{k\kappa}\over 8}^{1/2}}
e^{-{\kappa^{1/2}\over 8}\vert n-n'\vert^{\gamma}}. $$
Let $T$ be the operator with $\la n\vert T\vert m \ra =e^{-{\kappa^{1/2}\over 8}\vert
n-m\vert^{\gamma}}$. Then we can make $\Vert\left( (V_{\Lambda_L}^{\lambda})^{-1}
M_{\Lambda_L}^0\right)^k\Vert\le {1\over 2^k}\Vert T\Vert$ by making $C_0(\gamma ) L^u
e^{-{\kappa^{1/2}\over 8}}<\half$. We can therefore iterate the resolvent identity to get,
\begin{eqnarray*}
\Gamma^{\lambda}(0)&=&(V^{\lambda})^{-1}-(V^{\lambda})^{-1}M^0\Gamma^{\lambda}(0)\nonumber\\
&=&\sum_{k=0}^{\infty}(-1)^k \left((V^{\lambda})^{-1}M^0\right)^k (V^{\lambda})^{-1}
\end{eqnarray*}
Hence, if we take $L>Q_3$ with $Q_3$ larger than $Q_1$ and $Q_2$ and sufficiently large that
$\half L^{2u}>8\ln (2C_0(\gamma )L^u)$, the result follows from Lemmas 4.1 and 4.2.\qed

\paragraph
Let $\beta$ be fixed as in Theorem 3.4 and $\kappa>\pi /2$. To prove $(RA)$ of (P1) and
condition (P2) we need to look at two regimes, $\vert\lambda\vert\ge e^{-L^\beta/2}$ and
$\vert\lambda\vert <e^{-L^\beta/2}$. The next lemma deals with the first regime, and the
Lemmas 4.6 - 4.9 with the second.

{\bf Lemma 4.5}: {\sl If $\vert\lambda\vert\ge e^{-L^\beta/2}$, $L>1$, $E\in\RR$ and $\epsilon
>0$, then
\begin{equation}
\PP ( d(E, \sigma (M^{\lambda}_{\Lambda_L})) <\epsilon ) <8\rho_b R^2L^2 e^{L^\beta/2}
\epsilon .
\end{equation}
} {\bf Proof}: First we need to find a bound on the density of the diagonal terms of
$M^{\lambda}$.
\begin{eqnarray}\label{opb:densitybound}
\sup_x \lim_{\epsilon\rightarrow 0}{1\over {2\epsilon}} \int_{x-\epsilon <1-{\lambda\over
\omega}<x+\epsilon}\rho (\omega )d\omega &=&\sup_x \lim_{\epsilon\rightarrow 0}{1\over
{2\epsilon\lambda}} \int_{x-\epsilon}^{x+\epsilon} \rho\left( {\lambda\over
{1-u}}\right)\left( {\lambda\over {1-u}}\right)^2 du\nonumber\\ &<&\sup_x
\lim_{\epsilon\rightarrow 0} {{\rho_b R^2}\over {2\epsilon\vert\lambda\vert}}
\int_{x-\epsilon}^{x+\epsilon}du ={{\rho_b R^2}\over \vert\lambda\vert}.
\end{eqnarray}
It follows that the density of $x_{nn}=\la n\vert M_{\Lambda}^{\lambda}\vert n\ra$ is bounded
by $\rho_b R^2 e^{L^\beta/2}$.

For Borel subsets $B$ of $\RR$ let $\sigma^\Lambda_n(B) = \la n \vert E_\Lambda (B) \vert n
\ra$, where $ E_\Lambda (B) $ are the spectral projections of $ M^{\lambda}_\Lambda $. Then by
Lemma VIII.1.8 in Ref. 14, and by (\ref{opb:densitybound}) $$ \EE_{x_{nn}} \sigma^\Lambda_n
(B) < \rho_b R^2 e^{L^\beta/2} \int_B dx $$ and therefore $$ \EE \sigma^\Lambda_n (B) < \rho_b
R^2 e^{L^\beta/2}\int_B dx. $$
As in Proposition VIII.4.11 of Ref. 14, it then follows that, using (\ref{opb:densitybound}),
for all $E\in\RR$ and $\epsilon >0$, $$ \PP ( d(E, \sigma (M^{\lambda}_{\Lambda_L})) <\epsilon
) <2\rho_b R^2 e^{L^\beta/2} \epsilon \vert \Lambda_L \vert <2\rho_b R^2 e^{L^\beta/2}
\epsilon (L+1)^2. $$
If $L\ge 1$ the result follows.\qed

For the next part it is necessary to make some definitions.

{\bf Definitions:} We define $V=\{\zeta_n\, , n\in\ZZ [i]\}$ to be the vertices of a graph
${\cal G}(V,E)$ with edges $E$ defined as $E=\{ (\zeta_m,\zeta_n)\ :\
\vert\zeta_m-\zeta_n\vert <1/8,\ \forall\ n,m \in \ZZ [i], n\ne m \}$. The {\sl degree} of a
vertex, ${\hbox {\rm deg}}(\zeta_m)=\#\{n\in\ZZ [i]:\ (\zeta_m,\zeta_n)\in E\}$. Two vertices
are said to be {\sl connected} if there exists a path between them along a series of edges. A
{\sl component} is defined to be a maximally connected subgraph. We will define a {\sl
cluster} to be a component of the graph ${\cal G}(V,E)$.

\par {\bf Lemma 4.6}: {\sl For each configuration $\{\zeta_n\}$ there exist clusters
containing at most 4 vertices such that the distance between every pair of clusters is at
least $1/8$.}

{\bf Proof}: The distance between two clusters ${\cal C}_i$, ${\cal C}_j$ is given by,
$$ d({\cal C}_i,{\cal C}_j)=\inf\{\, d(\zeta_n^i,\zeta_m^j)\,\vert\,\zeta_n^i\in {\cal
C}_i,\,\zeta_m^j\in {\cal C}_j\}. $$
It is easily seen that if the distance between two clusters is less than $1/8$, then the
distance between one of the vertices in one cluster, must be closer than $1/8$ to a vertex in
the other. Thus an edge will exist that connects the two clusters, leading to a contradiction
in their definition as separate clusters.

It suffices to show that we cannot have a cluster with more than four vertices. The diameter
of a cluster is given by, $$ {\hbox {\rm diam}}({\cal C}_i)=\sup\{\, d(\zeta_n,\zeta_m)\
\vert\ \zeta_n,\zeta_m\in {\cal C}_i\}. $$ We know that the unit squares centred on the
Gaussian integers,  $\{B_n, n\in\ZZ [i]\}$, contain exactly one vertex each. The maximum
diameter for a cluster of five vertices will be less than 1/2 by virtue of the definition of a
cluster. However a circle of diameter 1/2 cannot intersect more than four of the $B_n$, so we
cannot have a cluster of five. If we had a cluster of more than five vertices, we could also
have a cluster of five as can be seen if we perform a one by one deletion of lowest degree
vertices until only five remain. Thus we cannot have a cluster with more than five
vertices.\qed

For a configuration $\{\zeta_n\}$, let $$ \langle n\vert {\tilde M}^{\lambda}_{\Lambda}\vert
n'\rangle = \cases{\langle n\vert M^{\lambda}_{\Lambda}\vert n'\rangle  & if $\zeta_n
,\zeta_{n'}$ are in the same cluster, \cr {0} & otherwise. \cr} $$ Let ${\cal C}_1, {\cal
C}_2,\ldots ,{\cal C}_N$ be the clusters in $\Lambda$ and let ${\cal P}_1, {\cal P}_2, \ldots
,{\cal P}_N$ be the projections onto ${\cal H}_i$ the space spanned by $\{\vert n\ra\,
:\,\zeta_n\in {\cal C}_i\}$. Let
\begin{equation}\label{opb:M_i}
M_i={\cal P}_i M^{\lambda}_{\Lambda} {\cal P}_i ={\cal P}_i {\tilde M}^{\lambda}_{\Lambda}
{\cal P}_i.
\end{equation}

{\bf Lemma 4.7}: {\sl For $\lambda =0$,
\begin{equation}
\vert\det M_i\vert\ge\quad C\prod_{m<\, n :\, \zeta_m ,\,\zeta_n\in\, {\cal C}_i}
\(1-e^{-\kappa\vert\zeta_m -\zeta_n\vert^2}\),
\end{equation}
where $C>0$ is a constant independent of $\zeta$. }

Note that numerical calculation shows that the inequality is satisfied with $C=1$. If the
$\zeta_i$'s are distinct then for $\lambda =0$ we can write
\begin{eqnarray}\label{lemm_ineq}
\la \xi, M_i\xi\ra &=&{\pi\over 2\kappa}\sum_{m,n:\zeta_m,\zeta_n\in {\cal C}_i}{\bar \xi}_m
P_0(\zeta_m,\zeta_n)\xi_n={\pi\over 2\kappa}\int_{\CC}\sum_{m,n:\zeta_m,\zeta_n\in {\cal
C}_i}{\bar \xi}_m P_0(\zeta_m,z)P_0(z,\zeta_n)\xi_n\ dz\nonumber\\ &=&
\int_{\CC}\vert\sum_{m:\zeta_m\in {\cal C}_i}\xi_m f_{\zeta_m}(z)\vert^2dz\ >\ 0
\end{eqnarray}
since the $f_{\zeta_i}$'s will be linearly independent. Thus $\det M_i>0$.

{\bf Proof:} If $\lambda =0$, recall that from the definition of $M^{\lambda}$, for $\zeta_m$,
$\zeta_n$ in a cluster ${\cal C}_i$, $$ \langle m\vert M_i\vert n\rangle
=e^{-\kappa\vert\zeta_m-\zeta_n\vert^2-2i\kappa\zeta_m \wedge\zeta_n}. $$ We only have to
prove the result up to a cluster of four. For a cluster of one, the result is trivial. For a
cluster of two we get, $$ \vert\det M_i\vert =1-e^{-2\kappa\vert\zeta_1-\zeta_2\vert^2}. $$
We now give the proof for a cluster of three. A direct proof with $C=1$ can be given (see Ref
16) but it is difficult to extend this to the case of a cluster of four impurities. For this
reason we shall give an indirect proof which can be extended to the latter case.

Let $\kappa^{\half}(\zeta_2-\zeta_1)=a e^{i\alpha}$ and $\kappa^{\half}(\zeta_3-\zeta_1) =b
e^{i\beta}$. Then $\det M_i = G_3(a,b,\phi)$ where $\phi =\alpha -\beta$ and
\begin{equation}\label{main1}
G_3(a,b,\phi)=1 - e^{-2a^2}- e^{-2b^2}-e^{-2c^2}+2e^{-(a^2+b^2+c^2)}\cos(2ab\sin (\phi)),
\end{equation}
with
$$ c^2=a^2+b^2-2ab\cos\phi. $$
Note that without loss of generality we can take $\phi\in [0,\pi ]$.  $G_3$ is an analytic
function of $a$, $b$ and $\phi$. It is easy to check that,
$$ G_3(0,b,\phi)=G_3(a,0,\phi)=G_3(a,ae^{\pm i\phi},\phi)=0, $$ and
$$ \frac{\partial G_3} {\partial a }(0,b,\phi) =\frac{\partial G_3} {\partial b }(a,0,\phi)=0
$$
so that we can write
\begin{equation}\label{factor1}
G_3(a,b,\phi)=a^2b^2(b-ae^{-i\phi})(b-ae^{i\phi})g_3(a,b,\phi)=a^2b^2c^2g_3(a,b,\phi)
\end{equation}
where $g_3(a,b,\phi)$ is an analytic function of $a$, $b$ and $\phi$.

Let $A=\dot{\RR}_+^2\times [0,\pi ]$, where $\dot{\RR}_+$ denotes the one-point
compactification of $\RR_+$, and let $A_o$ be the interior of $A$. Define
$f_3:A_o\rightarrow\RR$ by
\begin{equation}\label{total1}
f_3(a,b,\phi)={G_3(a,b,\phi)\over (1-e^{-a^2})(1-e^{-b^2})(1-e^{-c^2})}.
\end{equation}
$f_3(a,b,\phi)>0$ for all $(a,b,\phi )\in A_o$ by the inequality (\ref{lemm_ineq}). Note that
$c=0$ only if $a=b=0$ or $a=b$ and $\phi =0$. We shall prove that for each point
$(a_0,b_0,\phi_0)$ on the boundary of $A$, we have
$\displaystyle{\liminf_{(a,b,\phi)\rightarrow (a_0,b_0,\phi_0)}}$ $f_3(a,b,\phi)>0$. Then
since $A$ is compact there exists $C>0$ such that $f_3(a,b,\phi)>C$ for all $(a,b,\phi)\in A$.

For points on the boundary of $A$ for which $a$, $b$ and $c$ are all finite and non-zero
$f_3(a,b,\phi )$ is defined by (\ref{total1}) and is strictly positive. Now
$\lim_{a\rightarrow\infty}f_3(a,b,\phi)=1+e^{-b^2}>1$ for all $(b,\phi )\in\RR_+\times [0,\pi
]$ and similarly for $\lim_{b\rightarrow\infty}f_3(a,b,\phi)$. Also
$\displaystyle{\liminf_{(a,b)\rightarrow(\infty,\infty )}}$ $f_3(a,b,\phi)=1$.

Next we have that $\lim_{(a,b)\rightarrow (0,0)}$ $f_3(a,b,\phi)=g_3(0,0,\phi)$ and we can
check that $g_3(0,0,\phi )=4$. For $b>0$, $\lim_{a\rightarrow 0}
f_3(a,b,\phi)=b^4g_3(0,b,\phi)/(1-e^{-b^2})^2$. We can calculate $g_3(0,b,\phi )$ explicitly
to get $b^4g_3(0,b,\phi)=2e^{-2b^2}(e^{2b^2}-1-2b^2)>0$. Similarly we can show that
$g_3(a,0,\phi )>0$. Finally, by symmetry it follows that $\lim_{a\rightarrow b,\
\phi\rightarrow 0}f_3(a,b,\phi)=\lim_{a\rightarrow 0}f_3(a,b,\psi)$ where $\psi$ is the angle
between the edges of lengths $b$ and $c$, which has already been shown to be strictly
positive. Note that in fact this limit is independent of $\psi$.

Now we come to the proof of the Lemma for a cluster of four. The idea of the proof is the same
as for a cluster of three but the details are more complicated.

Let $\kappa^{\half}(\zeta_2-\zeta_1)=a e^{i\alpha}$, $\kappa^{\half}(\zeta_3-\zeta_1)=b
e^{i\beta}$ and $\kappa^{\half}(\zeta_4-\zeta_1)=c e^{i\gamma}$. Then $\det M_i =
G_4(a,b,c,\phi ,\psi)$ where $\phi =\beta -\alpha $, $\psi =\alpha -\gamma$ and
$$ G_4(a,b,c,\phi,\psi)=1 - e^{-2a^2}- e^{-2b^2}-e^{-2c^2} -e^{-2u^2} -e^{-2v^2}-e^{-2w^2} $$
$$
 + e^{-2(b^2+v^2)}+e^{-2(a^2+w^2)}+e^{-2(c^2+u^2)}
$$ $$ +2e^{-(a^2+b^2+u^2)}\cos(4\Delta_{abu}) +2e^{-(b^2+c^2+w^2)}\cos(4\Delta_{bcw}) $$ $$
+2e^{-(a^2+c^2+v^2)}\cos(4\Delta_{acv}) +2e^{-(u^2+v^2+w^2)}\cos(4\Delta_{uvw}) $$ $$
-2e^{-(b^2+c^2+u^2+v^2)}\cos(4(\Delta_{acv}+\Delta_{abu})) $$ $$ -2e^{-(a^2+
c^2+u^2+w^2)}\cos(4(\Delta_{abu}-\Delta_{bcw})) $$
\begin{equation}\label{main2}
-2e^{-(a^2+b^2+v^2+w^2)}\cos(4(\Delta_{acv}-\Delta_{bcw}))
\end{equation}
with
$$ u^2=b^2+a^2-2ba\cos\phi\ , $$ $$ v^2=c^2+a^2-2ac\cos\psi\ , $$ $$
w^2=b^2+c^2-2bc\cos(\phi+\psi)\ , $$ $$\Delta_{abu}=\half ba\sin\phi\ , $$ $$
\Delta_{acv}=\half ac\sin\psi\ , $$ $$ \Delta_{bcw}=\half bc\sin(\phi+\psi)\ , $$ and $$
\Delta_{uvw}=\half\left( ba\sin\phi+ac\sin\psi-bc\sin(\phi+\psi)\right)\ . $$
$G_4(a,b,c,\phi,\psi)$ is an analytic function of $ a,b,c $, $\phi$ and $\psi$. In this case
also we can check that, $$ G_4(0,b,c,\phi,\psi)=G_4(a,0,c,\phi,\psi)=G_4(a,b,0,\phi,\psi)=0,
$$ $$ \frac{\partial G_4} {\partial a }(0,b,c,\phi,\psi) =\frac{\partial G_4} {\partial b
}(a,0,c,\phi,\psi) =\frac{\partial G_4} {\partial c }(a,b,0,\phi,\psi)=0, $$ and $$
G_4(be^{\pm i\phi},b,c,\phi,\psi)=G_4(ce^{\pm i \psi},b,c,\phi,\psi) =G_4(a,ce^{\pm
i(\phi+\psi)}\,c,\phi,\psi)=0. $$ These identities imply that
\begin{equation}\label{factor2}
G_4(a,b,c,\phi,\psi)=a^2b^2c^2u^2v^2w^2 g(a,b,c,\phi,\psi)
\end{equation}
where $ g_4(a,b,c,\phi,\psi)$ is an analytic function of $ a,b,c $, $\phi$ and $\psi$.

In this case we let $A=\dot{\RR}_+^3\times [0,\pi ]^2$ and define $f_4:A_o\rightarrow \RR$ by
\begin{equation}\label{total2}
f_4(a,b,c,\phi,\psi)={G_4(a,b,c,\phi,\psi)\over
(1-e^{-a^2})(1-e^{-b^2})(1-e^{-c^2})(1-e^{-u^2})(1-e^{-v^2})(1-e^{-w^2})}.
\end{equation}
Using the same arguments as before it is sufficient to check that for each point
$z=(a_0,b_0,c_0,\phi_0,\psi_0)$ on the boundary of $A$, we have
$\displaystyle{\liminf_{\smash{(a,b,c,\phi,\psi)\rightarrow z}}}\, f_4(a,b,c,\phi,\psi)>0$.

When one of $a$, $b$ and $c$ tend to $\infty$, the problem simplifies to the three impurity
case and taking the lower limit when two of them tend to $\infty$, reduces the problem to the
two impurity case. When all of $a$, $b$ and $c$ tend to $\infty$ the lower limit is equal to
1. It remains to show that $f_4(a,b,c,\phi ,\psi )$ is strictly positive in the limit of any
subset of $\{a,b,c\}$ going to zero. By symmetry we need only check the cases
$a,b,c\rightarrow 0$, $a,b\rightarrow 0$ and $a\rightarrow 0$. Now $\lim_{(a,b,c)\rightarrow
(0,0,0)}$ $f_4(a,b,c,\phi,\psi)=g(0,0,0,\phi,\psi)=16/3$. Similarly $\lim_{(a,b)\rightarrow
(0,0)}$ $f_4(a,b,c,\phi,\psi)=$ $4e^{-2c^2}(e^{2c^2}-1-2c^2-2c^4) /(1-e^{-c^2})^3>0$.

Finally we need to check that $\lim_{a\rightarrow 0}f_4(a,b,c,\phi,\psi)>0$. This is
considerably more difficult and will be checked over several stages. We have that
\begin{equation}\label{expr}
\lim_{a\rightarrow 0}f_4(a,b,c,\phi,\psi)=2h(b,c,\theta)/(1-e^{-b^2})^2 (1-e^{-c^2})^2
(1-e^{-w^2})
\end{equation}
where
\begin{eqnarray}\label{h}
h(b,c,\theta)=&&1-(1+2b^2)e^{-2b^2}-(1+2c^2)e^{-2c^2}-e^{-2w^2}\nonumber\\ &&
+2w^2e^{-2(b^2+c^2)}+2e^{-b^2+c^2+w^2}\cos (2bc\sin\theta)\nonumber\\ && +
4bce^{-(b^2+c^2+w^2)}(\cos\theta \cos (2bc\sin\theta )+\sin\theta \sin (2bc\sin\theta ))
\end{eqnarray}
and $\theta =\phi+\psi$. Now differentiating $h$ with respect to $\theta$ gives
\begin{equation}\label{d_theta}
{dh\over d\theta}= 8bce^{-(b^2+c^2+w^2)}S( b c ,\theta)
\end{equation}
where $S(x,\theta)=\sin\theta\left( \cosh (2x\cos\theta )-\cos (2x\sin\theta )\right) -x\sin
(2x\sin\theta )$.

We can use $\cosh t\ge 1+t^2/2!+t^4/4!$ for all $t$ and $\sin t\le\sum_{n=0}^4
(-1)^nt^{2n+1}/(2n+1)!$, $\cos t\le \sum_{n=0}^4 (-1)^n t^{2n}/(2n)!$ for $t < 10$ to write
\begin{equation}
S(x,\theta)\ge {2\over 45}x^4 j(x,\theta)\sin\theta
\end{equation}
where $j(x,\theta)=15+2x^2\sin^6\theta -6x^2\sin^4\theta+4/7x^4\sin^6\theta-1/7x^4\sin^8\theta
-2/63x^6\sin^8\theta$. Note that $j(x,\theta)$ is symmetric about $\pi/2$. Differentiating
$j(x,\theta)$ with respect to $\theta$ yields $4/7x^2\sin^3\theta\cos\theta (21\sin^2\theta
-42+6x^2\sin^2\theta-2x^2\sin^4\theta-4/9x^4\sin^4\theta )$. The first factor is seen to be
zero only at $\theta\in\{ 0,\pi/2 ,\pi\}$. The second factor is quadratic in $x^2$ and has no
real roots as $(3-\sin^2\theta)^2+28/3(\sin^2\theta-2)<0$ for all $\theta$. Hence
$j(x,\theta)$ takes stationary values for $\theta\in\{ 0,\pi/2 ,\pi\}$. Now
$j(x,0)=j(x,\pi)=15$, while
$$ j(x,\pi/2)=15-4x^2+{3\over 7}x^4-{2\over 63}x^6 $$ is a cubic in $x^2$ which is easily
shown to be decreasing with $x$. $j(2.4,\pi/2)>0$ and thus for $x<2.4$, ${dh\over d\theta}\ge
0$ and $h$ is increasing with $\theta$.

We now need to find an increasing lower bound for $h$ when $x\ge 2.4$. For $\theta\in
[0,\pi/2]$, so that $\cos\theta\ge 0$, we can get a lower bound for $h$ by using $\cos
(2bc\sin\theta )\ge 1-2b^2c^2\sin^2\theta$ and $\sin (2bc\sin\theta )\ge -2bc\sin\theta$. Let
this lower bound be $k_1$. Note that $h$ and $k_1$ coincide at $\theta =0$. We have that
\begin{equation}\label{k_1}
{dk_1\over d\theta}=16bc\sin\theta e^{-(b^2+c^2+w^2)}S_1(b c,b c \cos\theta)
\end{equation}
where $S_1(x,y)=\sinh^2 y -2y+(2+y)(x^2-y^2)-y^2$. For $x\ge y$ we have that
$S_1(x,y)\ge\sinh^2 y-2y-y^2>0$ for $y>1.65$. On the other hand $S_1(x,y)>-2y+2(x^2-y^2)>0$ if
$y<(-1+\sqrt{1+4x^2})/2$. Combining these two results we have $S_1(x,y)\ge 0$ for all $y\le x$
if $x>2.1$.

For $\theta\in [\pi/2,\pi]$ we have that $\cos\theta\le 0$. We use the same bounds in $h$ as
before except we bound $\cos (2bc\sin\theta )$ by $1$ in the term containing $\cos\theta$. Let
this bound be $k_2$. Note that $k_1$ and $k_2$ coincide at $\theta =\pi/2$. It is simple to
see that
\begin{equation}\label{k_2}
{dk_2\over d\theta}=8bc\sin\theta e^{-(b^2+c^2+w^2)}S_2(b,c,\theta)
\end{equation}
where $S_2(b,c,\theta)=2\sinh^2(bc\cos\theta)-4bc\cos\theta+3b^2c^2\sin^2\theta\ge 0$. Hence
$k_2$ is increasing with $\theta$.

We have shown, when $x\ge 2.4$, that for $\theta\in [0,\pi/2]$, $k_1\le h$ is increasing, and
for $\theta\in [\pi/2,\pi]$, $k_2\le h$ is increasing. We have also noted that $h=k_1$ at
$\theta =0$ and that $k_1=k_2$ at $\theta =\pi/2$. We have seen that for $x<2.4$, $h$ is
increasing. Therefore it remains to check that $h$ is strictly positive at $\theta =0$. Making
the change of variables $s=2b^2$ and $r=2b(c-b)$ we have that
$$ h(\sqrt{s/2},(r/s+1)\sqrt{s/2},0)={s\over r^2}e^s e^{r^2/s}\ {\tilde h}(s,r) $$
where
\begin{eqnarray*}
{\tilde h}(s,r)&=&{(e^{r^2/s} - 1)\,s\,(e^{s} - 1 - s)\over r^2} + 2 {e^{ - r}\,s\,(1 - e^{-
r})\over r}\\  &&\mbox{\hspace{5.5truecm}} - e^{ - 2r}\,(1 - e^{- s}) - {s\,(1 + s)\,(1 - e^{-
r}) ^2\over r^2}.
\end{eqnarray*}
Let $k_3$ be the lower bound obtained by replacing the first term in ${\tilde h}$ by
$(1+r^2/2s)(e^s-1-s)$. We differentiate $k_3$ with respect to $r$ and write it as a power
series expansion in $r$;
\begin{eqnarray*}
s\,r^3\,e^{2r}\,{dk_3\over d r} &=& 2s(1-e^{-s}-s+{s^2\over 2})r^3+ (e^s-1-s-{s^2\over
2}+{5s^3\over 6})r^4\\ & & \mbox{\hspace{40pt}} +2s^2\sum_{n=5}^{\infty}{r^n\over
n!}\left(s(2^n-n-2)+2^n-n^2-n-2)\right)\\ & & \mbox{\hspace{60pt}} +(e^s-1-s)\sum_{n=5}^{\infty}{{2^{n-4}r^n}\over {(n-4)!}}.
\end{eqnarray*}
By using the bound $1-e^{-s}\le s$ it is easy to see that the first term is increasing and
thus positive. $e^s-1-s-{s^2\over 2}+{5s^3\over 6}\geq 0$ and $s(2^n-n-2)+2^n-n^2-n-2\geq 0$ for
$n\geq 5$. Thus $k_3$ is increasing with $r$.

Finally we need to check that the $\lim_{r\rightarrow 0}k_3(s,r)> 0$.
$$ \lim_{r\rightarrow 0}k_3(s,r)=2(\cosh s-1-{s^2\over 2})>0 $$
for $s>0$ and therefore $h(b,c,0)>0$ for all non-zero $b$, $c$ and the Lemma is proved.\qed

Recalling (\ref{opb:M_i}) we can write,
$$ {\tilde M}^{\lambda}_{\Lambda} =\sum_i M_i\qquad {\hbox { and }}\qquad ({\tilde
M}^{\lambda}_{\Lambda})^{-1} =\sum_i (M_i)^{-1}. $$
Let ${\displaystyle {\cal E}= \sup_i\Vert M_i^{- 1}\Vert }$, then $\Vert ({\tilde
M}^{\lambda}_{\Lambda})^{-1}\Vert\le {\cal E}$.

{\bf Lemma 4.8:}  {\sl Let $\delta_\kappa =\half e^{-\kappa /64}$. If $\theta>39$ and
$q<\theta /13 -3$ then there exists $Q_4$ such that for all $L>Q_4$, $\kappa >64\ln (2L^\theta
)$, for all $E\in (-\delta_\kappa ,\delta_\kappa )$, all $\lambda$ with $\vert\lambda\vert
<e^{-L^\beta/2}$,
$$ \PP\left( d(E,\sigma (M_{\Lambda_L}^{\lambda}))<e^{-L^\beta}\right) <{1\over L^q}. $$ }

{\bf Proof:} If $d_i =\dim {\cal P}_i$ then we have
\begin{equation}
\Vert M_i^{-1}\Vert\le{{c_{d_i}}\over {\vert\det M _i\vert }}\Vert M_i\Vert^{d_i-1}
\end{equation}
where $c_{d_i}$ is a constant. Obviously $d_i\in \{ 1,2,3,4 \}$, as a maximal cluster contains
4 impurities. Now, from the previous lemma we have a lower bound for $\vert\det M_i\vert$ for
the case where $\lambda =0$. Using the bound $1-\exp (-\kappa x^2)\ge x^2 \exp (-x^2)$ for
$\kappa>1$, we can write when $\lambda =0$,
\begin{equation}
\vert\det M_i\vert\ge\quad C\prod_{m<\, n :\, \zeta_m ,\,\zeta_n\in\, {\cal
C}_i}\vert\zeta_m-\zeta_n\vert^2 e^{-\vert\zeta_m-\zeta_n\vert^2}.
\end{equation}
Let $u=\theta/13$. Then if $\vert\zeta_n-\zeta_{n'}\vert>2/L^u$ for $\zeta_n,\zeta_{n'}\in
{\cal C}_i$ with $\zeta_n\ne\zeta_{n'}$, $$ \vert\det M_i\vert\ge C\({4e^{-(3/8)^2}\over
L^{2u}}\)^6=C'L^{-12u}. $$
Also, if $\lambda =0$ then $$ \Vert M_i\Vert^{d_i-1}\le\left( \sum_{\{n,m:\zeta_n,\zeta_m\in
{\cal C}_i\} } \vert\langle m\vert M_i\vert n\rangle \vert \right)^{d_i-1} <A, $$
for some constant $A$, independent of $\zeta$. So for $\lambda=0$, $\Vert M_i^{-1}\Vert\le
C''L^{12u}$.

Therefore if $D$ is the diagonal matrix made up of the elements $\lambda\over\omega_n$ with
$\vert {\lambda\over\omega_n}\vert<e^{-L^\beta/4}$ for $\{n:\zeta_n\in{\cal C}_i\}$ then for
$L$ sufficiently large, by the resolvent identity
\begin{equation}
{\cal E}=\sup_i\Vert M_i^{-1}\Vert\le \sup_i {\Vert M_i^{-1}\vert_{\lambda =0}\Vert\over
1-\Vert D\Vert\,\Vert M_i^{-1}\vert_{\lambda =0}\Vert}\le L^\theta.
\end{equation}
The probability for this to occur is greater than $\PP (\vert\omega_n\vert > e^{-L^{\beta}/4}
{\hbox { and }} \vert\zeta_n -\zeta_{n'}\vert>{2\over L^u}$ for all $n,n' {\hbox { such that
}}  \zeta_n ,\zeta_{n'}\in{\cal C}_i$ with  $\zeta_n\ne\zeta_{n'})$ which is greater than
$(1-2\rho_be^{-L^{\beta}/4})^4(1-L^{3-u}) >1-L^{-q}$ for $L$ sufficiently large.

Let $\delta M^{\lambda}_{\Lambda}=M^{\lambda}_{\Lambda} -{\tilde M}^{\lambda}_{\Lambda}$. Then
$$ \langle n\vert \delta M^{\lambda}_{\Lambda}\vert n'\rangle = \cases{ 0 & if $\zeta_n
,\zeta_{n'}$ are in the same cluster, \cr {\pi\over {2\kappa }}P_0(\zeta_n,\zeta_{n'} ) &
otherwise. \cr} $$
Since $\kappa\ge\pi /2$ we have, $$ \Vert\delta M^{\lambda}_{\Lambda}\Vert\le
e^{-{\kappa/64}}. $$
>From the resolvent identity we get
\begin{equation}
\Vert ( M^{\lambda}_{\Lambda})^{-1}\Vert\le \Vert ({\tilde M}^{\lambda}_{\Lambda})^{-1}\Vert
+\Vert ({\tilde M}^{\lambda}_{\Lambda})^{-1}\Vert\, \Vert \delta M^{\lambda}_{\Lambda} \Vert\,
\Vert ( M^{\lambda}_{\Lambda})^{-1}\Vert.
\end{equation}
If we can make $\Vert ({\tilde M}^{\lambda}_{\Lambda})^{-1}\Vert\, \Vert \delta
M^{\lambda}_{\Lambda}\Vert\le {\cal E}e^{-\kappa /64} \le\half$ then we get,
\begin{equation}
\Vert ( M^{\lambda}_{\Lambda})^{-1}\Vert\le 2\Vert ( {\tilde
M}^{\lambda}_{\Lambda})^{-1}\Vert.
\end{equation}
Thus, we have that if $\kappa >64\ln (2L^\theta )$, $$ \Vert (
M^{\lambda}_{\Lambda})^{-1}\Vert\le 2L^\theta, $$ with a probability greater than $1-L^{-q}$
if $L$ is sufficiently large.

If $\vert E\vert <\half e^{-\kappa /64}$ then $\vert E\vert <1/4L^{\theta}$. So $\Vert
(M_{\Lambda}^{\lambda}-E)^{- 1}\Vert <[\Vert
(M_{\Lambda}^{\lambda})^{-1}\Vert^{-1}-1/4L^\theta ]^{-1}.$ Hence from above, $$ \PP\left(
\Vert ( M^{\lambda}_{\Lambda}-E)^{-1}\Vert\le 4L^\theta \right) >1-{1\over L^q}. $$ Now, as
$E\in\RR$, $$ d(E,\sigma (M_{\Lambda}^{\lambda}))=\Vert ( M^{\lambda}_{\Lambda}-E)^{-1}
\Vert^{-1}, $$ which gives us that $$ \PP\left( d(E,\sigma (M_{\Lambda}^{\lambda})) \le
{1\over {4L^\theta }}\right) <{1\over L^q}. $$
Taking $Q_4$ sufficiently large so that in addition $L^\beta >\ln (4L^\theta)$ we obtain the
result.\qed

{\bf Lemma 4.9}: {\sl There exists $Q_5$ such that for all $L>Q_5$, for $q>0$, any $E\in\RR$,
all $\kappa <L^\beta/20$, all $\lambda$ with $\vert\lambda\vert < e^{-L^\beta/2 }$,
\begin{equation}
\PP ( d(E, \sigma (M^{\lambda}_{\Lambda_L})) <e^{-L^\beta} ) <{1\over L^q}.
\end{equation}
}

{\bf Proof}: We divide up the points of $\Lambda\cap\ZZ [i]$ into adjacent pairs $\{
n_i,n_i^\prime \}$.

Let $Q_i$ be the two dimensional projection onto the space spanned by $\vert n_i\rangle$ and
$\vert n_i^\prime \rangle$. Let
$$ U_i={1\over \sqrt{2}}\left(\matrix{ 1&-e^{- 2i\kappa\zeta_{n_i}\wedge\zeta_{n_i^\prime}}\cr
1&e^{- 2i\kappa\zeta_{n_i}\wedge\zeta_{n_i^\prime}}\cr}\right)\  ,\quad U_i^* ={1\over
\sqrt{2}}\left(\matrix{ 1&1\cr -e^{ 2i\kappa\zeta_{n_i}\wedge\zeta_{n_i^\prime}}&e^{
2i\kappa\zeta_{n_i}\wedge\zeta_{n_i^\prime}}\cr}\right)\qquad $$
Let $U=\sum_i Q_i U_i Q_i^*$. Then we have, for $n_i, n_i^\prime$ in a pair,
\begin{eqnarray*}
\langle n_i\vert UM^0 U^*\vert n_i\rangle &=&1+e^{-\kappa\vert\zeta_{n_i}-
\zeta_{n_i^\prime}\vert^2}\\ \langle n_i^\prime\vert UM^0 U^*\vert n_i^\prime\rangle
&=&1-e^{-\kappa\vert \zeta_{n_i}- \zeta_{n_i^\prime}\vert^2}
\end{eqnarray*}
Now, if $r=\vert\zeta_{n_i}-\zeta_{n^\prime_i}\vert$, then
\begin{eqnarray*}
\PP [r\in (a,b)] &=&\int d^2\zeta_1\int d^2\zeta_2 r(\zeta_1)\, r(\zeta_2)\,1_{\{\vert\zeta_1
-\zeta_2\vert\in (a,b)\} }\\ &\le &\int r(\zeta_1 )d^2(\zeta_1)\int_{\RR^2} r(\zeta_2)1_{
\{\vert\zeta_1 -\zeta_2\vert\in (a,b)\} }d^2(\zeta_2)\nonumber\\ &\le &2\pi r_b\int_a^b r dr
=\pi r_b (b^2 -a^2) =\int_a^b {\tilde \rho}(r)dr
\end{eqnarray*}
with ${\tilde \rho}(r)=2\pi r_b\, r$.

Let $s=e^{-\kappa r^2}$ and $e^{-5\kappa}<a<b<1$. Then $-(\pi r_b\, ds)/(\kappa s)={\tilde
\rho}(r)dr$
$$ \PP [s\in (a,b)]< -{{\pi r_b}\over\kappa}\int_a^b {ds \over s} <{{\pi r_b}\over\kappa}
e^{5\kappa}\int_a^b ds={{\pi r_b}\over\kappa} e^{5\kappa}(b-a). $$
The density of $x_{nn}=\langle n\vert UM^0 U^*\vert n\rangle$ is bounded by $\pi
r_be^{5\kappa}/\kappa$. So the diagonal terms $1\pm s$ have the same bound.

For Borel subsets $B$ of $\RR$ let $\sigma^\Lambda_n(B) = \la n \vert E_\Lambda (B) \vert n
\ra$, where $ E_\Lambda (B) $ are the spectral projections of $ UM^0 U^*$, with
$\vert\lambda\vert < e^{-L^\beta/2}$. Then by Lemma VIII.1.8 in Ref. 14,
$$ \EE_{x_{nn}} \sigma^\Lambda_n (B) < {{\pi r_be^{5\kappa}}\over\kappa} \int_B dx $$
and therefore
$$ \EE \sigma^\Lambda_n (B) < {{\pi r_be^{5\kappa}}\over\kappa}\int_B dx. $$
As in Proposition VIII.4.11 of Ref. 14, it then follows that for all $E\in\RR$ and $\epsilon
>0$,
\begin{equation}
\PP ( d(E, \sigma (M^0_{\Lambda_L})) <\epsilon ) <2{{\pi r_be^{5\kappa}}\over\kappa} \epsilon
\vert \Lambda_L \vert <2{{\pi r_be^{5\kappa}}\over\kappa} \epsilon (L+1)^2 <8{{\pi
r_be^{5\kappa}}\over\kappa}L^2 \epsilon.
\end{equation}
if $L\ge 1$. Now, it is easily seen that $\sigma (M^\lambda )\subset \{\, z\, :d(z,\sigma(M^0
))<\Vert M^\lambda -M^0\Vert\}$. Hence
$$ d(E, \sigma (M^\lambda ))\ge d(E, \sigma (M^0)) - \Vert M^\lambda -M^0  \Vert. $$
We can show that $\Vert M^\lambda -M^0  \Vert <e^{-L^\beta /4}$ with a probability $\PP
(\vert\omega_n\vert > e^{-L^\beta/4}\ \ {\hbox { for all }} n\in\Lambda_L\ )
>(1-2\rho_be^{-L^\beta/4})^{(L+1)^2} >1-L^{-2q}$ for all
$L>Q_5$ with $Q_5$ sufficiently large. Hence if $d(E, \sigma (M^0_{\Lambda_L}))>\epsilon
+e^{-L^\beta/4}$ then $d(E,\sigma (M^\lambda_{\Lambda_L} ))>\epsilon$.

So for $\lambda$ with $\vert\lambda\vert < e^{-L^\beta/2 }$, all $E\in\RR$ and for all
$L>Q_5$,
\begin{equation}
\PP ( d(E, \sigma (M^{\lambda}_{\Lambda_L})) >\epsilon )
>\left( 1- 8{\pi\over\kappa}r_b L^2 (\epsilon+e^{-L^\beta/4}
)e^{5\kappa}\right)\left( 1- {1\over L^{2q}}\right).
\end{equation}
where we have used $\PP (A)\ge \PP (A\vert B)\PP (B)$. If we have that $L^\beta>20\kappa$ the
lemma is proved.\qed

\par Finally, we can bring the two regimes for $\lambda$ together to prove that:

{\bf Lemma 4.10}: {\sl There exists $Q_6$ and $\delta_\kappa$ such that for any $q>0$, any
$\lambda\in \RR$, $E\in (-\delta_\kappa ,\delta_\kappa )$, for all $L>Q_6$, $$ \PP(d(E,\sigma
(M^\lambda_{\Lambda_L(0)}))<e^{-L^\beta})<{1\over L^q}. $$ }

{\bf Proof}: Choose $\theta>13(q+3)$. Take $Q_6$ larger than $Q_5$ such that
$$ 8\rho_b R^2 L^2e^{-L^\beta/2}<{1\over L^q},\quad {\hbox { and }}\quad L^\beta>2^8.5\ln
(2L^\theta ) $$
for all $L>Q_6$ and take $\delta_\kappa =\half e^{-\kappa /64}$. Let $E\in (-\delta_\kappa
,\delta_\kappa )$ and $\vert\lambda\vert<e^{-L^\beta/2}$. Then by Lemmas 4.8 and 4.9, for all
$L>Q_6$,
$$ \PP(d(E,\sigma (M^\lambda_{\Lambda_L(0)}))<e^{-L^\beta})<{1\over L^q}. $$ On the other hand
if $\vert\lambda\vert\ge e^{-L^\beta/2}$ then by Lemma 4.5, for all $L>Q_6$ we also have the
above inequality.\qed

We finally check that the conditions of Theorem 3.4 are satisfied for $p=3$, $q=25$. By Lemma
4.10 condition (P2) is satisfied for $L>Q_6$ with $\eta =\delta_{\kappa}$ where
$\delta_{\kappa}=\half e^{-\kappa /64}$. Also from Lemma 4.10 $(RA)$ of (P1) is satisfied with
probability greater than $1-{1\over L^q}$ for $L>Q_6$.

Now in Lemma 4.4 put $u=7$, $\gamma =\half$ and let $L_0$ be greater than $Q_3\vee Q_6$ and
such that for any fixed $s\in (\half ,1)$ (as in the regularity condition) we have $$ {1\over
L_0^{25}}+{2\over L_0^4}<{1\over L_0^3},\qquad L_0^{14}(L_0-L_0^s)^{1/2}>64\ln (2L_0^7),\quad
{\hbox {\rm and }}\quad L_0^7(L_0-L_0^s)^{1/2}>32L_0. $$ If we choose $\kappa_0=L_0^{4u}/4$
then we have that
$$ \vert\la n\vert \Gamma_{\Lambda_{L_0}}^\lambda (0)\vert n'\ra\vert\le e^{-\kappa^{1/4}L_0}
$$ for all $\kappa >\kappa_0$, all $n$, $n'\in\Lambda_{L_0}$ with a probability greater than
$1-2/L_0^4$. Therefore if we take $m_0=\kappa^{1/4}$,
$$ \PP\{\Lambda_{L_0}(0){\hbox { is }}(m_0 ,0)-{\hbox {regular }}\}\ge 1-{1\over L_0^3} $$
where we have used that $\PP (A\cap B)\ge 1-\PP (A^c)-\PP (B^c)$ and condition (P1) is
checked.

\section{Proof of Theorem 3.2 parts (b) and (c).}

In this section we denote by ${\cal L}$ the Lebesgue measure.

>From Theorem 3.4, equation 3.5 and an application of Fubini's Theorem we can deduce that with
probability one and for ${\cal L}$-a.e. $\lambda$,\ if $\lambda$ is a non-zero generalized
eigenvalue of $H(\omega,\zeta)$ then the corresponding eigenfunction decays exponentially. An
immediate consequence is that $\sigma_{\hbox {\rm ac }}(H)=\emptyset$. However this does not
rule out the existence of singular continuous spectrum. To exclude this we need to show
exponential decay for a.e. $\lambda$ with respect to the spectral measure of
$H(\omega,\zeta)$. We will use the ideas of Delyon, L\'evy and Souillard$^{17}$.

Let $\Lambda\subset\ZZ [i]$ with $\vert\Lambda\vert =N$ as before and define the restriction
of $H$ to $\Lambda$ by,
$$ H_{\Lambda} =\sum_{n\in\Lambda} \omega_n\vert f_{\zeta_n}\rangle\langle f_{\zeta_n}\vert.
$$
If $\psi_k$ are eigenfunctions of $H_{\Lambda}$ with eigenvalues $\lambda_k$, $k=1,\ldots ,N$
and $\Vert \psi_k\Vert =1$ for all $k$, then $$ H_{\Lambda}\psi_k =\lambda_k \psi_k. $$
Define the resolution of the identity of the restriction of $H$ to $H_{\Lambda}$ by
$$ E_{\Lambda}(B)=\sum_{k:\lambda_k\in B} \vert \psi_k\rangle\langle\psi_k\vert\ , $$
where $B$ is a Borel subset of $\RR$ and let $\sigma_{\Lambda}^{\phi ,\phi }=\langle\phi\vert
E_{\Lambda}\vert\phi\rangle$ for some $\phi\in {\cal H}_0$. For $\Lambda\nearrow \ZZ [i]$,
$\sigma_{\Lambda}^{\phi ,\phi}$ converges weakly as a measure to $\sigma^{\phi ,\phi}$, the
spectral measure of $H$. We can write $\sigma_{\Lambda}^{\phi ,\phi}$ explicitly:
\begin{equation}\label{opb:specmeas}
\sigma_{\Lambda}^{\phi ,\phi} (B)=\sum_{k:\lambda_k\in
B}\vert\langle\phi\vert\psi_k\rangle\vert^2\,.
\end{equation}
As in Section III the eigenvalues $\lambda_k$ of $H_\Lambda$ with eigenfunction $\psi_k$ must
satisfy $M_\Lambda^\lambda\xi_k =0$ where $\xi_k (n)=\sqrt {2\kappa /\pi}\ \omega_n\la
f_{\zeta_n}\vert \psi_k\ra$. Thus we can expect $N$ solutions of $\lambda$ for $\det
M_\Lambda^{\lambda }=0$. We will look at solutions as a function of one of the $\omega_n$
only.

Using ${{\partial H_{\Lambda} }\over {\partial\omega_n }}=\vert f_{\zeta_n} \rangle\langle
f_{\zeta_n} \vert$, a calculation of $\langle\psi_k\vert d/d\omega_n (H_{\Lambda}\psi_k)
\rangle$ yields that
\begin{equation}\label{opb:dlambdadomega}
{{d\lambda_k }\over {d\omega_n }}=\vert\langle f_{\zeta_n} \vert \psi_k\rangle\vert^2\,.
\end{equation}
Note that if $\langle f_{\zeta_n}\vert\psi_k\rangle=0$ then $\psi_k$ remains an eigenvector
for $\lambda_k$ as $\omega_n$ varies and does not contribute to $\sigma_{\Lambda}^{f_{\zeta_n}
,f_{\zeta_n}}$. Also, if $\lambda_k$ is degenerate, we can choose the corresponding
orthonormal set of eigenvectors so that only one satisfies $\langle
f_{\zeta_n}\vert\psi_k\rangle\ne 0$. From (\ref{opb:dlambdadomega}) we see that each
$\lambda_k$ is a monotonic increasing function of $\omega_n$ and from (3.4) that we get $N-1$
solutions of $\lambda_k$ which are identical as $\omega_n\to\pm\infty$. The $N$-th solution
corresponds to the $\psi_k$ which tends to $f_{\zeta_n}$ and this value increases from
$\lambda =-\infty$ at $\omega_n=-\infty$ to the lowest $\lambda_k$ at $\omega_n=+\infty$
(respectively increases from the highest $\lambda_k$ at $\omega_n=-\infty$ to $\lambda
=+\infty$ at $\omega_n=+\infty$) as can be seen from the following argument. We would like to
know the behaviour when $\omega_n$ and $\lambda$ both become large. If we expand the
determinant we get,
$$ \(1-{\lambda\over {\omega_n}}\)\({{(-1)^{N-1}}\over\Pi}\lambda^{N-1}
+P(\lambda)\)+\sum_{m\ne n}^N \omega_m {{e^{-2\kappa\vert\zeta_m -\zeta_n\vert^2}
(-1)^{N-2}}\over \Pi}\lambda^{N-2}+Q(\lambda)=0 $$
where $\Pi =\prod_{m\ne n}^N\omega_m$, $P(\lambda)$ is a polynomial in $\lambda$ of degree
$N-2$ and $Q(\lambda)$ is a polynomial in $\lambda$ of degree $N-3$. We thus get an expression
for $\omega_n$ in terms of $\lambda$.
$$ \omega_n ={{\lambda^N}\over {\lambda^{N-1}-\sum_{m\ne n}^N \omega_me^{-2\kappa\vert\zeta_m
-\zeta_n\vert^2}\lambda^{N-2} +R(\lambda )}} $$
where $R(\lambda )$ is a polynomial in $\lambda^{N-3}$. Thus for $\lambda$ large, $\omega_n
\sim\lambda$. This is what we expect if $\psi_k\to f_{\zeta_n}$ as then we have that
$d\lambda_k /d\omega_n \to 1$.

Now recalling the fact that $\lambda $ is a monotonic increasing function of $\omega_n$ we see
that only one among the eigenvalues $\lambda_k$ crosses any $\lambda$, i.e. the range of
$\lambda (\omega_n )$ is divided into disjoint open subsets $O_k$ such that $\bigcup_k O_k
=\RR$ and each $\lambda_k$ corresponds to only one of the $O_k$. Therefore there is only one
term corresponding to such $\lambda_k$ in the sum (\ref{opb:specmeas}) for
$\sigma_{\Lambda}^{f_{\zeta_n} ,f_{\zeta_n}}$.

The above results along with (\ref{opb:specmeas}) and (\ref{opb:dlambdadomega}) allow us to
make the following change of variables:
\begin{eqnarray}
\int_{\RR\times B} \rho(\omega_n )d\omega_n\sigma_\Lambda^{f_{\zeta_n}
,f_{\zeta_n}}(d\lambda)&=&\int_{\RR}\sum_{k:\lambda_k\in B}\rho
(\omega_n)d\omega_n\vert\langle f_{\zeta_n}\vert\psi_k\rangle\vert^2\nonumber\\ &\le &\rho_b\
\#\{\lambda_k\in B\}=\rho_b\,\vert B\vert .
\end{eqnarray}
Using the weak convergence of $\sigma_\Lambda^{f_{\zeta_n},f_{\zeta_n}}$ to $\sigma_{\phantom
\Lambda}^{f_{\zeta_n},f_{\zeta_n}}$ we can therefore write
\begin{equation}\label{opb:boundavspecmeas}
\int_{\RR}\rho(\omega_n )d\omega_n\sigma^{f_{\zeta_n} ,f_{\zeta_n}}(B )\le\rho_b\ \vert B\vert
,
\end{equation}
and hence the $\omega_n$-averaged spectral measure $\EE_{\omega_n}\(\sigma^{f_{\zeta_n}
,f_{\zeta_n}}(d\lambda )\)$ is absolutely continuous with respect to Lebesgue measure.

We now use Kotani's ``trick'' (see Ref. 12). In the following, ${\cal B}$ will represent the
Borel $\sigma$-field. We will need the following lemma who's proof is elementary.

{\bf Lemma 5.1 :}\ {\sl Let $\{ f_n \}$ be a total countable subset of normalised vectors of a
Hilbert space ${\cal H}$ and $H$ a self-adjoint operator on ${\cal H}$ with spectral
projections $E(\,\cdot\, )$. Let $c_n > 0$, $\sum_n c_n <\infty$ and $\nu =\sum_n c_n
\sigma^{f_n,f_n}$, where $\sigma^{f_n,f_n}(\cdot) =\langle f_n\vert E(\cdot )\vert
f_n\rangle$. Then  for any $B\in {\cal B}$, $\nu (B)=0$ implies that $E(B)=0$.}

Let $(\Omega, {\cal F}, \PP )$ be the probability space corresponding to the $\omega$ and
$\zeta$ and let ${\cal F}_n^*$ be the sub $\sigma$-field of ${\cal F}$ generated by all of
these variables except $\omega_n$ for some $n\in\ZZ [i]$. If $F(\omega ,\zeta ,\lambda )$ is a
nonnegative ${\cal F}^*_n\otimes {\cal B}$ measurable function, then from Proposition VIII.1.4
in Ref. 14 we have that,
\begin{equation}\label{opb:kotani}
\(\EE\{\int F(\,\cdot\, ,\,\cdot\, ,\lambda )\,d\lambda\} =0\)\Rightarrow\(\int F(\omega
,\zeta ,\lambda )\sigma^{f_{\zeta_n},f_{\zeta_n}}(d\lambda )=0\ \PP {\hbox {\rm -a.e. }}\).
\end{equation}

{\bf Lemma 5.2 :}\ {\sl For $B\in {\cal B}$ let $B\mapsto E (B)$ be the spectral measure of
$H$ and let $A\in \cap_{n\in\ZZ [i]}({\cal F}^*_n\otimes {\cal B})$, then, if for a.e.
$\lambda\in \RR$ with respect to Lebesgue measure $\EE\{ 1_A (\,\cdot ,\,\cdot ,\lambda )\}
=0$, then $\EE\{ E(\{\lambda :\, (\,\cdot ,\,\cdot ,\lambda )\in A\} )\} =0$. }

{\bf Proof :}\ Let $A\in {\cal F}^*_n\otimes {\cal B}$. If for a.e. $\lambda$ with respect to
Lebesgue measure $\EE\{ 1_A\{(\,\cdot ,\,\cdot ,\lambda )\} =0$, then by Fubini's Theorem
$\EE\{\int d\lambda 1_A (\,\cdot ,\,\cdot ,\lambda )\} =0$. Combining this with
(\ref{opb:boundavspecmeas}) we have that $\EE\{\int 1_A (\,\cdot ,\,\cdot ,\lambda )
\EE_{\omega_n} (\sigma^{f_{\zeta_n},f_{\zeta_n}}(d\lambda))\} =0$. We now use the fact that
$A\in {\cal F}^*_n\otimes {\cal B}$ with (\ref{opb:kotani}) to move the expectation over
$\omega_n$ outside the integral and we obtain
\begin{equation}
\EE\{ 1_A\{(\,\cdot ,\,\cdot ,\lambda )\} =0\quad\Rightarrow\quad\EE\{\int 1_A (\,\cdot
,\,\cdot ,\lambda )\sigma^{f_{\zeta_n},f_{\zeta_n}}(d\lambda)\} =0.
\end{equation}
Finally, by taking $A\in \cap_{n\in\ZZ [i]}({\cal F}^*_n\otimes {\cal B})$ and $\nu =\sum_n
c_n \sigma^{f_{\zeta_n},f_{\zeta_n}}$, where each $c_n > 0$, $\ \sum_n c_n <\infty$ we have
that $\EE\{\int_{\RR} 1_A (\,\cdot ,\,\cdot ,\lambda )\nu (d\lambda )\}=0$ and from Lemma 5.1
we have the result. Now we have seen at the beginning of this section that if $W$ is the set
in $\Omega\times\RR$ defined by:
\begin{eqnarray*}
W=&&\{ (\omega ,\zeta ,\lambda ):\ {\hbox {\rm the  generalized eigenfunctions of }} H(\omega
,\zeta)\ \\ &&\hskip 6truecm{\hbox {\rm with eigenvalue }}\,\lambda {\hbox { decay
exponentially }}\},
\end{eqnarray*}
then Fubini's Theorem implies that $W$ is of $\PP\otimes {\cal  L}$ full measure. By taking
$W^c$ as $A$ in Lemma 5.2 we have shown that with probability one and $\lambda$-a.e. with
respect to the spectral measure, if $\lambda$ is a generalized eigenvalue of $H$ then the
corresponding eigenfunctions decay exponentially and hence Theorem 3.2 parts (b) and (c) are
proven.

\section*{Acknowledgements}

We would like to thank T. C. Dorlas for providing a direct proof of Lemma 4.7 in the case of
three impurities and for elucidating some aspects of Kotani's trick. One of us (M.S.)
acknowledges support from the Forbairt Scientific Research Scheme (Basic Research Scheme
SC/1997/621).
\newpage

\section*{References}

\noindent $^1$ J. Desbois, S. Ouvry, C. Texier, Nucl. Phys. B {\bf 500}, 486 (1997).

\noindent $^2$ T. C. Dorlas, N. Macris, J. V. Pul\'e, Helv. Phys. Acta {\bf 68}, 330 (1995);
J. Math. Phys. {\bf 37}, 1574 (1996).

\noindent $^3$ J. M. Combes, P. D. Hislop, Commun. Math. Phys. {\bf 177}, 603 (1996).

\noindent $^4$ W-M. Wang, J. Funct. Anal. {\bf 146}, 1 (1997).

\noindent $^5$ H. von Dreifus, A. Klein, Commun. Math. Phys. {\bf 124}, 285 (1989).

\noindent $^6$ J. Fr\"ohlich, T. Spencer, Commun. Math. Phys. {\bf 88}, 151 (1983).

\noindent $^7$ J. Bellisard, A. van Elst, H. Schulz-Baldes, J. Math. Phys. {\bf 35}, 5373
(1994).

\noindent $^8$ T. C. Dorlas, N. Macris, J. V. Pul\'e, J. Stat. Phys. {\bf 87}, 847 (1997).

\noindent $^9$ T. C. Dorlas, N. Macris, J. V. Pul\'e, Commun. Math. Phys. {\bf 204}, 367
(1999).

\noindent $^{10}$ M. Aizenman, S. Molchanov, Commun. Math. Phys. {\bf 157}, 245 (1993).

\noindent $^{11}$ M. Scrowston, Ph.D. thesis, National University of Ireland, Dublin (1999).

\noindent $^{12}$ S. Kotani: in {\sl Proceedings of the 1984 AMS conference on Random Matrices
and their Applications}, Contemp. Math. {\bf 50}, Providence RI (1986).

\noindent $^{13}$ J. V. Pul\'e, M. Scrowston, J. Math. Phys. {\bf 38}, 6304 (1997).

\noindent $^{14}$ R. Carmona, J. Lacroix: {\sl Spectral Theory of Random Schr\"odinger
Operators}. Birkh\"auser - Boston (1990).

\noindent $^{15}$ R. Ph. Boas: {\sl Entire Functions}. Academic Press - New York (1954).

\noindent $^{16}$ T. C. Dorlas: Private communication.

\noindent $^{17}$ F. Delyon, H. L\'evy, B. Souillard, Commun. Math. Phys. {\bf 100}, 463
(1985).

\end{document}